\documentclass{article}
\usepackage{color}

\usepackage{spconf,amsmath,graphicx}
\usepackage{balance}

\newcommand{\new}{\vskip 0.45em plus 0.15em minus .1em} %was {\vskip 0.75em plus 0.15em minus .1em}

\title{Simultaneously learning speaker\textquotesingle s direction and head orientation from binaural recordings}
\name{Harshvardhan Takawale, Nirupam Roy\vspace{-0.2in}}
\address{\vspace{-0.1in}University of Maryland College Park}
\begin{document}
% \newcommand{\niru}[1]{\textcolor{red}{#1}}
% \newcommand\red[1]{{\color{red}#1}}

%\ninept
%
\maketitle
\begin{abstract}
Estimation of a speaker's direction and head orientation with binaural recordings can be a critical piece of information in many real-world applications with emerging `earable' devices, including smart headphones and AR/VR headsets.
However, it requires predicting the mutual head orientations of both the speaker and the listener, which is challenging in practice.
This paper presents a system for jointly predicting speaker-listener head orientations by leveraging inherent human voice directivity and listener's head-related transfer function (HRTF) as perceived by the ear-mounted microphones on the listener.
We propose a convolution neural network model that, given binaural speech recording, can predict the orientation of both speaker and listener with respect to the line joining the two. 
The system builds on the core observation that the recordings from the left and right ears are differentially affected by the voice directivity as well as the HRTF. We also incorporate the fact that voice is more directional at higher frequencies compared to lower frequencies.
%Our system builds on the core observation that when sound travels from the speaker's mouth to both ears, it passes through different channels that depend on voice directivity and HRTF and thus contains cues about speaker and listener head orientations.
%This system of head orientation can be useful in many AR/VR and localization applications.
Our proposed system achieves $2.5^{\circ}$ 90th percentile error in the listener's head orientation and $12.5^{\circ}$ 90th percentile error for that of the speaker.
%Our evaluation shows that the system has $>$90\% accuracy in detecting listener head orientations with $<$7$^{\circ}$ error and $>$80\% accuracy in detecting speaker head orientations with $<$10$^{\circ}$ error.
\end{abstract}
% \begin{abstract} % Old version 1
% This paper discusses a system for estimating the mutual speaker-listener head orientation by leveraging inherent directivity in human speaking and listening. We develop a convolution neural network model that, given binaural speech recording, can predict the orientation of both speaker and listener with respect to the line joining the two. Our system builds on the core observation that when sound travels from the speaker's mouth to both ears, it passes through different channels that depend on voice directivity and HRTF and thus contains cues about speaker and listener head orientations. This system of head orientation can be useful in many AR/VR and localization applications. Our evaluation shows that the system has $>$90\% accuracy in detecting listener head orientations with $<$7$^{\circ}$ error and $>$80\% accuracy in detecting speaker head orientations with $<$10$^{\circ}$ error.
% \end{abstract}
%
\begin{keywords}
Voice directivity, HRTF, head orientation, voiced sounds, auditory perception
\end{keywords}
\section{Introduction}
\label{sec:intro}
\vspace{-0.1in}
It is long known that human voice and hearing are directional in nature \cite{dunn1939exploration,steven2006perceptual}. Thus, when humans speak or hear, the sounds contain cues specific to the orientation of the human head. This enables our brains the natural ability to localize sound sources and sense when they are being talked to. Thus, this paper aims to explore whether binaurally recorded human speech contains these cues necessary to estimate the head orientation of not just the listener but also the speaker. The intuition behind the idea is that as the sound travels through different channels before reaching each ear. So each of the recorded signals should encode some information regarding both the listener's and the speaker's head orientations. 

Our proposed system leverages the directivity of human voice and hearing at different frequencies. We show that these effects are enhanced in the near-field. The differences in interaural recordings are used to infer the orientations. As the human voice directivity pattern (VDP) is dynamic \cite{noufi2022reconstructing}, we first identify the segments that have higher directivity based on the harmonic structure \cite{10.1145/3539490.3539600} present in these regions and mask rest of the signal to suppress noise. We then extract the crucial features like interaural level difference (ILD), interaural time difference (ITD), and differences in the energy of frequency bins\cite{lindblom2014human} and use them as features to train a deep neural network.

We evaluate our system by creating a dataset that combines a real-world VDP dataset, a real-world HRTF dataset, and a real-world speech dataset. The evaluations show that the system has $2.5^{\circ}$ 90th percentile error in detecting listener head orientations and $12.5^{\circ}$ 90th percentile error in detecting speaker head orientations. Knowing these head orientations opens up context-aware applications in smart-home environments \cite{shen2020voice}, AR/VR \cite{binelli2018individualized}, and Human-robot interaction \cite{10.1145/2157689.2157834}.

% Contribution - 
% 1) Generating near-field synthetic dataset with voice directivity and hrtf 2) Using only 1 binaural recording to predict orientation of both listener and speaker 3) Use inverse convolution feature  to highlight difference in directivity of high and low frequency 4) show that voice directivity can be used to infer speaker facing direction 5) Show that near-field has more diversity and thus better for this application 6) Process ILD and ITD separately into low and high frequencies as head orientation affects them differnetly.

\begin{figure}[t]
    \centering
    \includegraphics[width=2.5in]{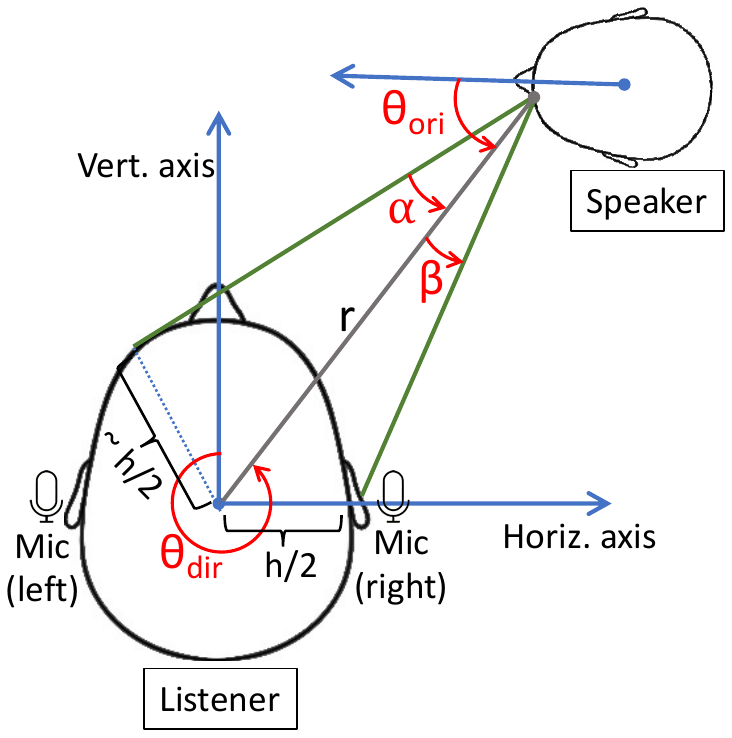}
\vspace{-0.16in}
\caption{Near field speaker-listener model overview.}
\vspace{-0.2in}
\label{fig:overview}
\end{figure}

\vspace{-0.15in}
\section{Problem Formulation and Intuitions}
\label{sec:format}
\vspace{-0.1in}

We aim to find the direction of a speaker as well as her head orientation on the horizontal plane in an egocentric reference frame centered on the listener.
The horizontal axis of this reference frame is along the line passing through the listener's ears and the perpendicular axis is along the listener's facing direction, as shown in Figure \ref{fig:overview}.
We now consider a line joining the centers of the listener and the speaker's heads.
The angle this line makes with the vertical axis is defined as the speaker's direction $\theta_{dir}$.
For the speaker's head orientation, we consider the angle $\theta_{ori}$ between the speaker's facing direction and the line joining the listener's center of the head and the speaker's mouth.
Note that $\theta_{dir}$ and $\theta_{ori}$ can vary independently of each other.
The goal is to estimate $\theta_{dir}$ and $\theta_{ori}$ simultaneously using only the binaural recording of natural speech from a pair of ear-mounted synchronized microphones on the listener.
%The direction of the speaker is defined as the positive angle $\theta_{dir}$ that the line joining the ce
%Assumptions: Speech signal, distance is within near field, two angles can move.
\new

\noindent{\bf Intuition:}
Humans have an excellent capability to perceptually find sounds' direction of arrival which is largely attributed to unique directional filtering by the listener's head, outer ears, and torso leading to differential binaural reception in the left and right ears.
This effect is known as Head-Related Transfer Function (HRTF) which leads to spatial hearing cues.
A large body of works leveraged HRTF to computationally estimate the sound's direction \cite{wang2020binaural,ben2018localization,mendoncca2012improvement}.
However, as a function of a person's individual physiological factors, the HRTFs do not generalize across users and it limits the achievable accuracy in direction estimation.
Binaural localization methods with body-worn microphones often assume prior knowledge of the listener's personal HRTF data to improve estimation accuracy \cite{oberem2020experiments}.
Moreover, existing work assumes HRTF is the only factor that alters the received sound ignoring the impacts from the orientation of the speaker's head.
%\new
%In our problem space, the source sound is human speech and it introduces a unique opportunity in estimating the orientation of the speaker.
%$\theta_{dir}$ and $\theta_{ori}$.
Existing measurements show that a human speaker does not radiate voice sounds uniformly in all directions, rather the sound shows a frequency-dependent radiation pattern due to the diffraction with the head.
This leads to the VDP that shows higher frequencies are attenuated more than lower frequency sounds toward the back of the head, which adds a cue for speaker's head orientation with respect to the listener.
Therefore, the binaural recording in the listener's ear-worn microphones manifests a combined effect of the speaker's VDP and listener's HRTF leading to a unique opportunity to estimate both the direction and orientation of the speaker ($\theta_{dir}$ and $\theta_{ori}$).
\new

\begin{figure}[h]
    \centering
    \includegraphics[width=1.1in]{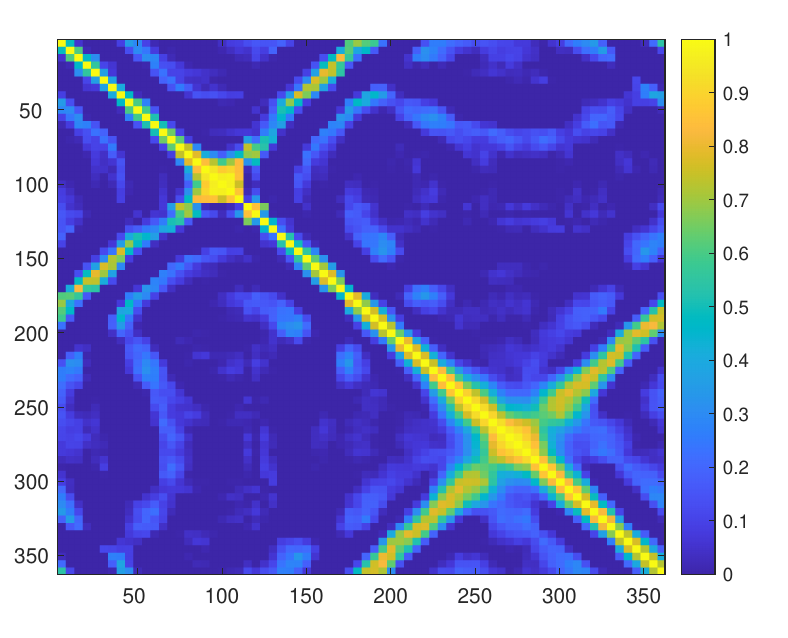}
    \includegraphics[width=1.1in]{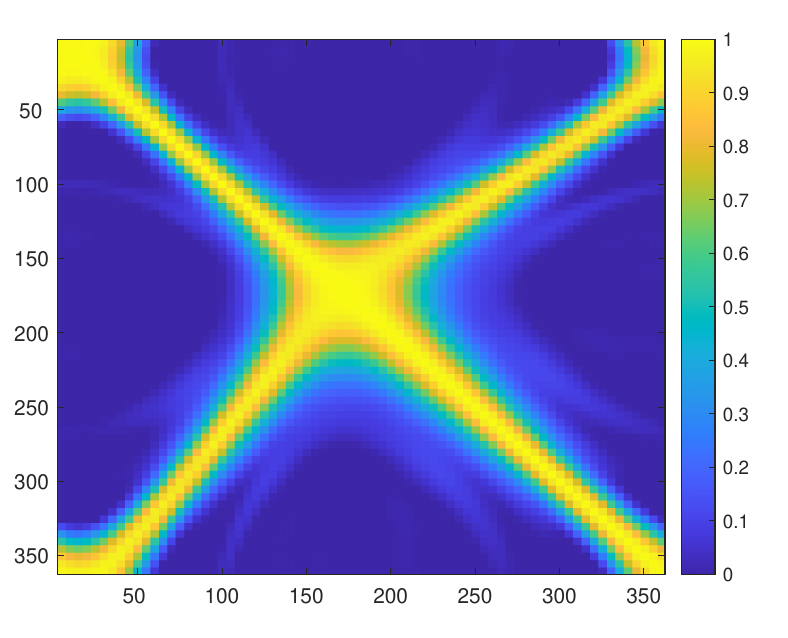}
    \includegraphics[width=1.1in]{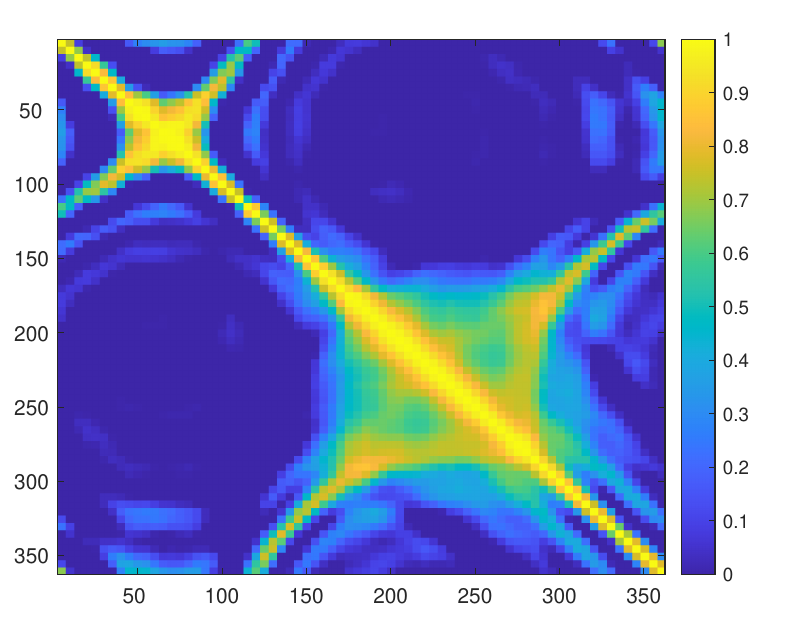}
\vspace{-0.2in}    
\caption{Correlation matrix of (a) HRTF, (b) VDP, and (c) combined HRTF and VDP.}
\vspace{-0.15in}    
\label{fig:hrtfcorr}
\end{figure}

Several recent works leveraged VDP to estimate the speaker's head orientation using standalone microphone arrays\cite{wei2021inferring}.
However, due to the lateral symmetry, the VDP-based head orientation estimation performs poorly with several tens of degrees of error \cite{arend2019synthesis,yang2021model}.
In this paper, we hypothesize that (a) it is possible to jointly estimate speakers' direction ($\theta_{dir}$) and orientation ($\theta_{ori}$) from a single binaural recording and (b) this joint parameter estimation can enhance the accuracy of both of these angles.
When we verify the diversity of HRTF and VDP across different angles separately using the correlation matrix shown in Figure \ref{fig:hrtfcorr}(a) and \ref{fig:hrtfcorr}(b), the VDP shows significant confusion.
However, a correlation matrix with combined features, as shown in Figure \ref{fig:hrtfcorr}(c), improves overall diversity indicating a possibility of joint estimation.
\new

%\vspace{-0.2in}
\vspace{-0.05in}

\noindent{\bf Challenges:}
We consider social conversational distance to be the physical scale where the speaker-listener orientation estimation can be applied.
Therefore we consider a near field region of sound for our model, which is between $0.5$ to $1.5$ meters in range.
However, a majority of the available HRTF datasets are collected at far-field.
Such datasets do not capture the substantial difference of interaural distances from the source in the near-field and also the diffraction and shadowing effects on nearby sound sources.
Moreover, the distance and angle between the centers of the speaker/listener heads differ significantly from those of ears in near-field.
This requires us to transform existing HRTF datasets for our near-field scenarios using a model elaborated next.

\new

\subsection{Near-field speaker-listener model:}
\vspace{-0.1in}
%We assume a pair of ear-mounted synchronized microphones on the listener produces binaural recordings of the sounds from the speaker.
%This microphone pair is the only source of signal to our system.
%We consider facing directions of a listener  in the horizontal plane.
%We define facing axis as the axis perpendicular to the line joining the ears and passing through the center of the head.
%We aim to estimate the facing axis makes with the line joining the centers of the speaker and listener's center of the head.
%We consider an egocentric listener with the earphones and a speaker. We fix the reference frame with respect to the listener, who facing the y-axis.
%We now consider a line joining the listener and the speaker and the angle between this line and the listener's facing direction is defined as $\theta_{dir}$ and the angle between the speaker's facing direction and the line is defined as $\theta_{ori}$.
When the speech signal originates from the mouth of the speaker to the left and right ears of the listener, the signal undergoes changes owing to the combination of head-related transfer function (HRTF) of the listener, $H(\theta_{dir})$,  and voice directivity pattern (VDP) of the speaker, $V(\theta_{ori})$. The left and right ears are separated by a distance of $h$ which is the width of the head. We define $r$ as the distance between the speaker and the listener. Thus, for the left and right ears, there is an additional offset in angle defined by $\alpha$ and $\beta$ respectively. When the right ear of the listener is ipsilateral to the speaker, these angles can be derived in terms of $h$, $r$, and $\theta_{dir}$ as follows. %(http://audiogroup.web.th-koeln.de/PUBLIKATIONEN/Arend_DAGA2019.pdf) - (this paper talks about the near-field corrections that the library does. Need to highlight that near field has head-shadowing effect and parallax effect that we are considering) - 
\vspace{-0.2in}
\begin{equation}
    \alpha = \arcsin(\frac{\frac{h}{2}}{r}), \ \ \beta = \arctan(\frac{\frac{h*cos\theta_{dir}}{2}}{r - \frac{h*sin\theta_{dir}}{2}})
\end{equation}
\vspace{-0.15in}
% \begin{equation}
%     \beta = \arctan(\frac{\frac{h*cos\theta_{dir}}{2}}{r - \frac{h*sin\theta_{dir}}{2}})
% \end{equation}

% \begin{figure}[t]
%     \centering
%     \centerline{\includegraphics[width=8.5cm]{Figures/overview2.jpg}}
% \caption{Problem overview}
% \label{fig:overview}
% \end{figure}

The resultant signal at the left and right ear in the frequency domain, $Y_{l}(f)$ and $Y_{r}(f)$ can thus be defined as follows. 
\begin{equation}
    \begin{array}{l}
%  \begin{align*}
    Y_{l}(f) = X(f)  H_{ln}  V(\theta_{ori} - \alpha)\\
    Y_{r}(f) = X(f)  H_{rn}  V(\theta_{ori} + \beta)
%  \end{align*}
    \end{array}
\end{equation}
% \begin{equation}
%     Y_{r}(f) = X(f)  H_{rn}  V(\theta_{ori} + \beta)
% \end{equation}

where $X(f)$ is the source speech signal in frequency domain and $H_{ln}$ and $H_{rn}$ are left and right HRTFs compensated for near field.

\begin{figure*}[t]
    \centering
    \centerline{\includegraphics[width=\linewidth, height=1.7in]{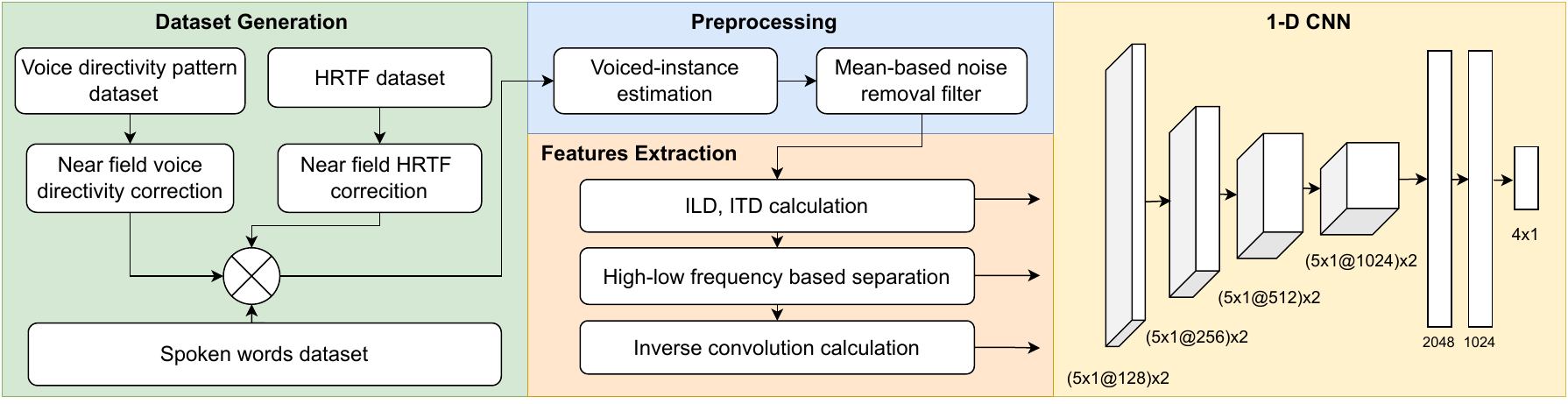}}
\vspace{-0.18in}
\caption{System overview}
\vspace{-0.2in}
\label{fig:dataset_generation}
\end{figure*}

\section{System Design}
\label{sec:system_design}
\vspace{-0.1in}
In this section, we elaborate on the inner workings of the system. The system can majorly be divided into 4 sections - 1) The dataset generation pipeline 2) the pre-processing pipeline 3) the Feature extraction pipeline and 4) the 1-D CNN-based regression model as shown in Figure \ref{fig:dataset_generation}.

\subsection{Binaural feature extraction:}
\vspace{-0.1in}
We extract the interaural level difference (ILD) and interaural time difference (ITD) to be used as core features.
We define ILD and ITD as follows.
\begin{equation}
    \begin{array}{l}
        ILD(\theta_{dir},\theta_{ori},f,t)  = 20\log_{10}(\frac{|Y_{l}(f,t)|}{|Y_{r}(f,t)|})\\
        ITD(\theta_{dir},\theta_{ori},f,t)  = \frac{1}{2\pi f}\angle(\frac{Y_{l}(f,t)}{Y_{r}(f,t)})                  
    \end{array} 
\end{equation}

% \begin{equation}
%     \begin{array}
%         ILD(\theta_{dir},\theta_{ori},f,t)  = 20\log_{10}(\frac{|Y_{l}(f,t)|}{|Y_{r}(f,t)|})\\
%                             % & \hspace{-0.4in}= 20\log_{10}(\frac{|H_{ln}(\theta_{dir}, f,t)V(\theta_{ori} - \alpha, f,t)|}{|H_{rn}(\theta_{dir}, f,t)V(\theta_{ori} + \beta, f,t)|})
%         ITD(\theta_{dir},\theta_{ori},f,t)  = \frac{1}{2\pi f}\angle(\frac{Y_{l}(f,t)}{Y_{r}(f,t)})                  
%     \end{array} 
% \end{equation}

%\noindent and inter ear time difference as:
% \begin{equation}
%     \begin{aligned}
%         ITD(\theta_{dir},\theta_{ori},f,t) & = \frac{1}{2\pi f}\angle(\frac{Y_{l}(f,t)}{Y_{r}(f,t)}) \\
%                              %& \hspace{-0.4in}= \frac{1}{2\pi f}\angle(\frac{H_{ln}(\theta_{dir}, f,t)V(\theta_{ori} - \alpha, f,t)}{H_{rn}(\theta_{dir}, f,t)V(\theta_{ori} + \beta, f,t)})
%     \end{aligned} 
% \end{equation}

\new

\subsection{Pre-processing for speech signal:}
\vspace{-0.08in}

% \subsection{Voice Directivity pattern}
% Past research has shown human speech has directional characteristics. The directivity is frequency-dependent. We perform a COMSOL simulation to show the propagation of sound coming out of the mouth. As seen in Fig \ref{fig:voice_directivity}, the radiation is omnidirectional for lower frequencies and its directivity increases with frequency. This is in accordance with previous real-world studies \cite{monson2012horizontal}. Thus, this voice directivity pattern contains cues that can help inferring the facing direction of the speaker, most importantly the relative energy difference for higher and lower frequencies. In section \ref{sec:inverse_convolution} we will discuss how we leverage this feature.

% \begin{figure}[h]
%     \centering
%     \centerline{\includegraphics[width=4cm]{Figures/voice-direcitvity-pattern.png}}
% \caption{Example of voice directivity pattern}
% \label{fig:voice_directivity}
% \end{figure}

\begin{figure}[h]
    \centering
    \centerline{\includegraphics[width=8.5cm, height=1.2in]{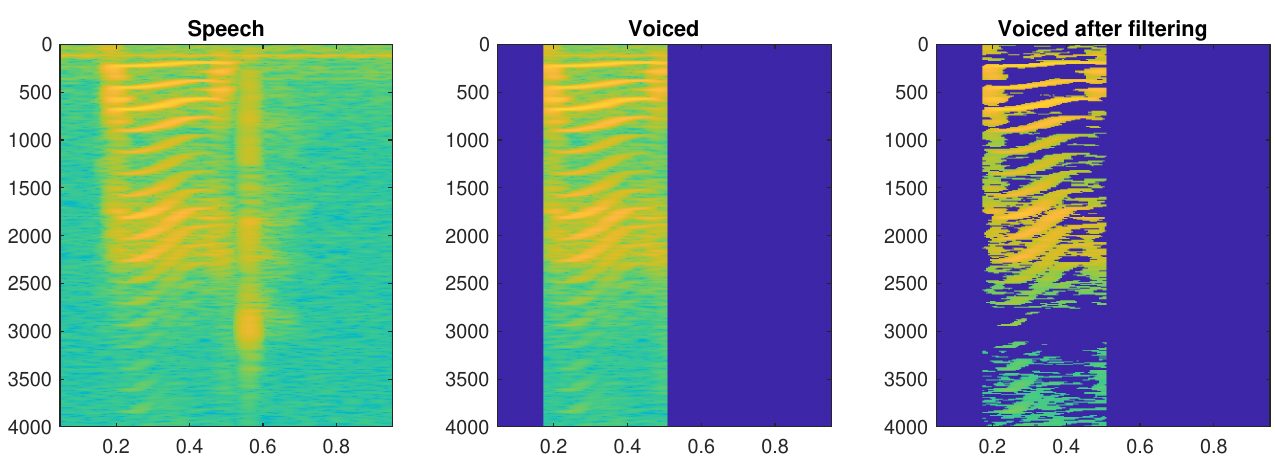}}
\vspace{-0.2in}    
\caption{Signal before and after voiced-unvoiced filtering}
\vspace{-0.3in}
\label{fig:voiced}
\end{figure}

\noindent{\bf Voiced vs Unvoiced speech: }
%As discussed in the previous section, 
Voiced speech has a higher directivity index. Thus, during preprocessing, we extract segments that contain voiced speech and mask the rest of the segments as shown in Fig \ref{fig:voiced}. We use the VOICEBOX toolbox on Matlab for this purpose. The toolbox uses PEFAC \cite{Gonzalez2014PEFACA} to first estimate the fundamental pitch and use it to find the probability of the segment being voiced.
1) We identify that the voiced section is sufficient to identify orientation as it has higher directivity. Thus we remove unvoiced sections that could otherwise introduce unnecessary noise. This is done by identifying the harmonic structure of the human speech.
\new

\noindent{\bf Missing frequencies in Human speech: }
As seen in Fig \ref{fig:voiced}, human speech has a harmonic structure. Thus, to reduce the noise in preprocessing, we zero force all the values below a threshold. The threshold is decided for each frequency bin based on the mean energy per bin.

\subsection{Near-field correction}
\vspace{-0.1in}

Most of the large real-world HRTF datasets are collected in the far field (e.g. RIEC dataset is collected at a distance of 1.5m). When trying to convert the far-field HRTF data, there are various near-field effects that need to be taken into account. As the source is nearby, the sound reaches the listener as a spherical wave instead of a planar wave and makes different angles with the ears. Also, the human head acts as a low-pass filter for the contralateral ear. To model these behaviors, \cite{arend2019synthesis} apply distance variation functions in the spherical harmonics domain. We follow the same method from the SUpDEq library. We also consider the parallax effect due to the directional nature of the speaker. We consider a simple spherical head model and speaker as a point source to calculate and adjust the VDP.

\vspace{-0.1in}

\subsection{Separating facing configurations in ILD-ITD plane}
\vspace{-0.1in}

To infer $\theta_{dir}$, we mainly rely on the ITD values. When the inter-time-difference value is less, it indicates that the listener is facing (towards or away from) the sound source. Having said that, the ITD values at higher frequencies undergo aliasing. The aliasing frequency depends on the size of listener's head $h$. if we consider $h$ = 18cm, the aliasing frequency would be 952 Hz based on the spatial aliasing limit. Thus we divide ITD values into two halves $ITD_{low}$ and $ITD_{high}$ before sending it to the machine learning model. To infer $\theta_{ori}$, we rely on the ILD values. As VDP has more directivity at higher frequencies, the $ILD$ values at higher frequencies, $ILD_{high}$, are affected more by the speaker-facing direction than the $ILD$ values at lower frequencies $ILD_{low}$. 

As $ILD_{low}$, $ILD_{high}$, $ITD_{low}$, and $ITD_{high}$ have their own unique characteristics, they are fed to the CNN as separate channels so that they can be treated as independent features.

% \begin{figure}[h]
%     \centering
%     \centerline{\includegraphics[width=8.5cm, height=2in]{Figures/ILD-ITD.png}}
% \caption{simplistic ILD ITD}
% \label{fig:overview}
% \end{figure}

% \begin{figure}[h]
%     \centering
%     \centerline{\includegraphics[width=8.5cm, height=2in]{Figures/voiced_mean_filtered.pdf}}
% \caption{Voiced signal after filtering}
% \label{fig:voiced_filtered}
% \end{figure}

%\subsection{High to low ratio (inverse convolution)}
\subsection{Enhancing VDF features with inverse convolution}
\label{sec:inverse_convolution}
\vspace{-0.1in}

As discussed earlier, voice radiation has different patterns for higher and lower frequencies. The higher frequencies have greater variation with direction \cite{dunn1939exploration}. On the other hand, lower frequencies are more omnidirectional. To emphasize this fact, we define a new feature - the ratio of energies of high and low frequencies - that can be useful to determine $\theta_{ori}$. We convolve $ILD_{high}$ with $1/ILD_{low}$ and use the resultant values as the fifth channel of input.

% Thus, a ratio of energies of high and low frequencies can be used as a feature to determine the speaker's orientation. Thus we divide the $ILD$ values into two parts - $ILD_{high}$ and $ILD_{low}$. We then convolve $ILD_{high}$ with $1/ILD_{low}$ and use the resultant values as the fifth channel of input.

\vspace{-0.1in}

\subsection{DNN model design}
\vspace{-0.08in}
%\vspace{-0.1in}
\noindent{\bf Architecture:}
We use a 1-D convolution neural network(CNN) with 8 convolution layers and 3 fully connected layers. We use dropout layer after the convolution layers for regularization. Each layer uses ReLU activation. The input contains 5 channels. We perform FFT on the 1-second recording of the dataset we created earlier and then pass it through the pre-processing pipeline. We use dropout regularization before the fully connected layers. The model is trained on NVIDIA Ampere A100 GPUs using ADAM optimizer with batch size 50 and learning rate $5 \times 10^{-4}$.
\new

% \noindent{\bf Input features}
% After we convert the binaural recording into frequency domain, we use the $ILD$, $ITD$, and inverse convolution values as the input features. We further divide the $ILD$ and $ITD$ values for higher and lower frequencies giving us $ILD_{high}$, $ILD_{low}$, $ITD_{high}$ and $ITD_{low}$ giving us 5 input channels.

% \begin{figure}[h]
%     \centering
%     \centerline{\includegraphics[width=8.5cm]{Figures/mlmodel.png}}
% \caption{ML model overview}
% \label{fig:ML_overview}
% \end{figure}

\vspace{-0.12in}
\section{Evaluation}
\label{sec:typestyle}
\vspace{-0.1in}

For generating data to determine listener and speaker head orientations, we use the following three datasets.
{\bf (a) HRTF dataset:} We use the RIEC database\cite{Kanji_Watanabe2014E1368}. The RIEC database contains HRTFs from 105 subjects (210 different ears). The database is measured in 5-degree intervals in azimuth and $10^{\circ}$ interval in elevation. For our study, we are only considering the azimuthal plane.
{\bf (b) VDP dataset:} We use the data from \cite{porschmann2022effects} \cite{porschmann2021investigating} and \cite{porschmann2023investigating} for the VDP. The database contains 288 VDP data. We only use the reference measurements of vowel utterances and ignore the measurements for other scenarios such as the subject holding a hand in front of the mouth or cupping the hands around the mouth.
{\bf (c) Spoken words dataset:} We use the Speech Commands database\cite{warden2018speech} for the spoken words. The database contains 105,829 utterances of 35 words from 2,618 subjects. 
%As shown in Fig \ref{fig:dataset_generation},
We generate the audio for our study we randomly pick an HRTF, a VDP, and a spoken word along with a random $\theta_{dir}$ value and $\theta_{ori}$ value. This gives us a diverse set of listeners, speakers, listener-facing direction, speaker-facing direction, and spoken words. Overall, we generated 560,000 recordings.
\vspace{-0.1in}

\subsection{Orientation Estimation}
\begin{figure}[h]
    \centering
    \includegraphics[width=1.67in]{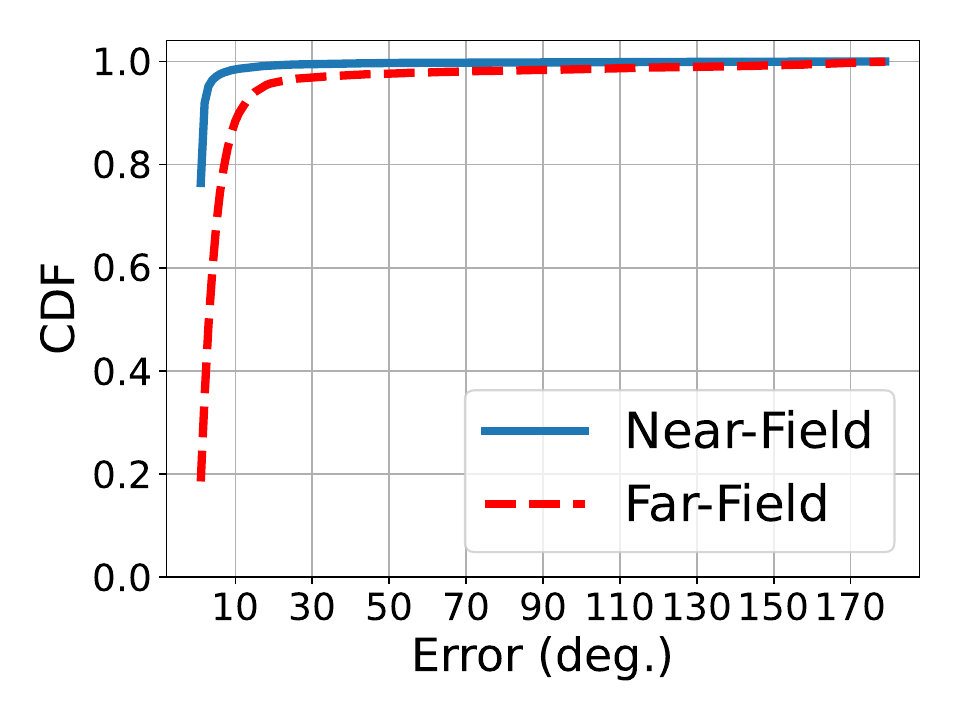}
    \includegraphics[width=1.67in]{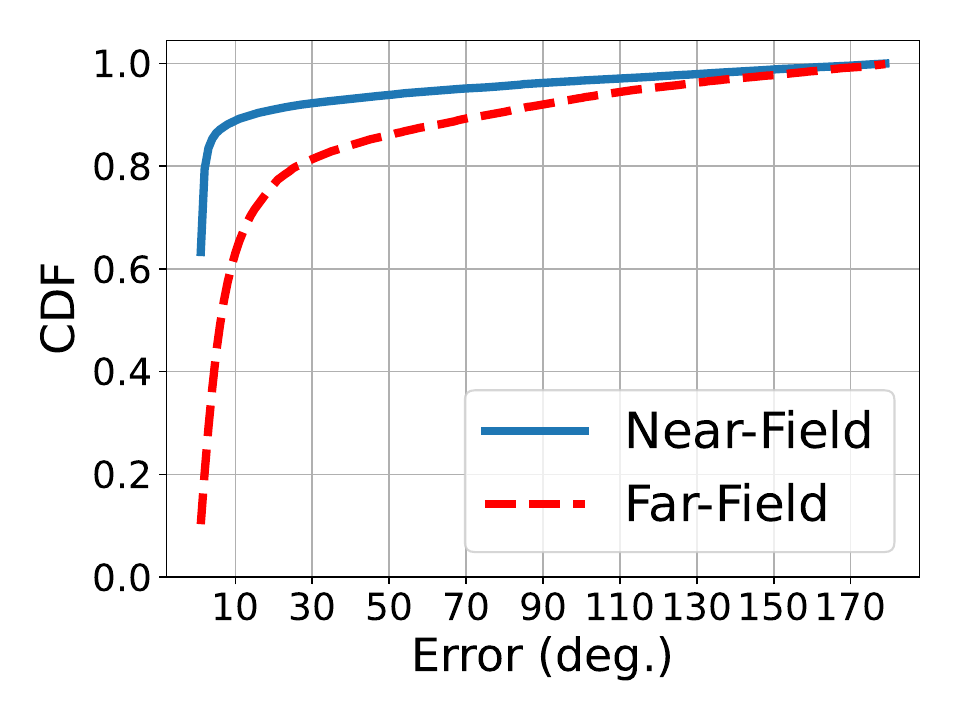}
\vspace{-0.4in}
\caption{CDF of error in (a) $\theta_{dir}$ and (b) $\theta_{ori}$ for near-/far-field.}
%\caption{Comparison CDF plot for error in estimated (a) listener and (b) speaker direction for near and far-field scenarios}
\vspace{-0.05in}
\label{fig:far_near}
\end{figure}

\vspace{-0.1in}
In this section, we discuss the accuracy of the model in predicting $\theta_{dir}$ and $\theta_{ori}$. We convert the ground-truth $\theta_{dir}$ and $\theta_{ori}$ into $sin(\theta_{dir})$, $sin(\theta_{ori})$, $cos(\theta_{dir})$ and $cos(\theta_{ori})$ and use mean-square error loss for training this model. In Fig \ref{fig:far_near} (a) the blue curve shows the cumulative distribution function (CDF) plot for the error in predicted listener head orientation. We reach the $90^{th}$ percentile for around $2.5^{\circ}$ error. In Fig \ref{fig:far_near} (b) the blue curve shows the CDF plot for error in predicted speaker head orientation with $90^{th}$ percentile error $12.5^{\circ}$.

%\vspace{-0.1in}

\begin{figure}[h]
    \centering
    \includegraphics[width=1.3in]{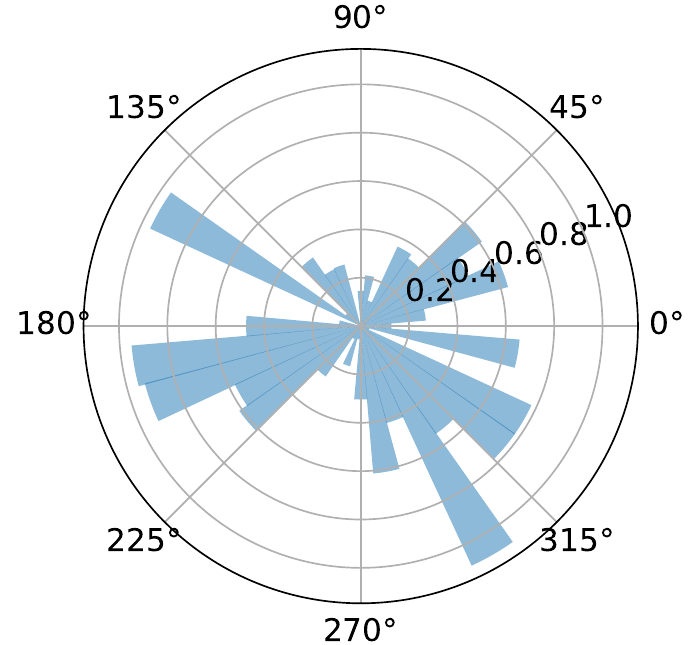}
    \includegraphics[width=1.3in]{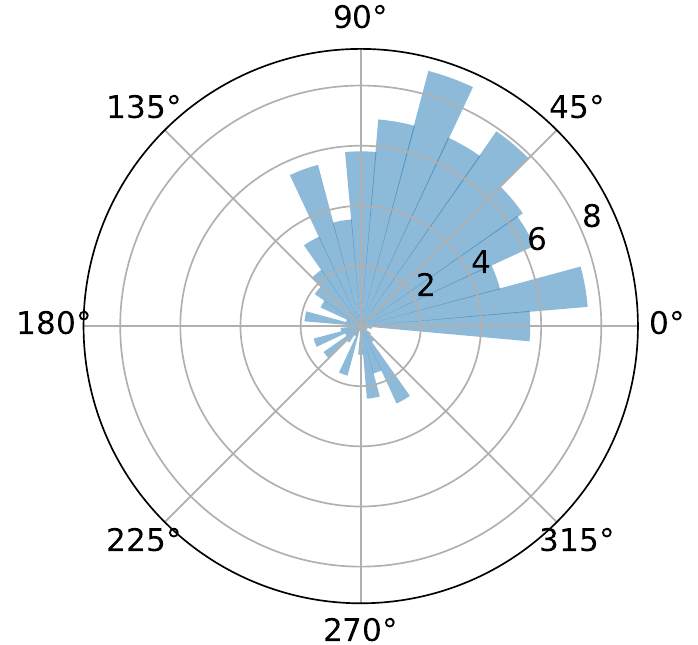}
\vspace{-0.17in}
\caption{Polar error plot of (a) $\theta_{dir}$ and (b) $\theta_{ori}$}
\label{fig:polar_all}
\end{figure}

To verify which azimuthal sector incurs a higher loss, we plot the error for each angular sector. Fig \ref{fig:polar_all} (a)
shows the error in $\theta_{dir}$ for every $<$10$^{\circ}$. The mean error in each sector is less than $<$1$^{\circ}$ for all sectors except one. Fig \ref{fig:polar_all} (b) shows the error in $\theta_{ori}$ for every $<$10$^{\circ}$. The mean error in each sector is less than $<$8$^{\circ}$ for the majority of the sectors. %all sectors except one.

% \begin{figure}[h]
%     \centering
%     \centerline{\includegraphics[width=8.5cm]{Figures/known_unknown.jpg}}
% \caption{Problem statement placeholder}
% \label{fig:overview}
% \end{figure}

% \begin{figure}[h]
%     \centering
%     \centerline{\includegraphics[width=8.5cm]{Figures/regression_cdf_speaker_known_unknown.pdf}}
% \caption{Comparison CDF plot for error in estimated speaker direction when HRTF is known vs. when HRTF is unknown}
% \label{fig:cdf_speaker_known_unknown}
% \end{figure}

%\vspace{-0.1in}

% \begin{figure}[h]
%     \centering
%     \centerline{\includegraphics[width=8.5cm]{Figures/polar_speaker.pdf}}
% \caption{Polar plot for error in estimated Speaker direction}
% \label{fig:polar_speaker}
% \end{figure}

\new
%\noindent{\bf How Near-field improves results:}
\noindent{\bf Impact of near-field correction:}
In this section, we evaluate how near-field aids in the estimation of $\theta_{dir}$ and $\theta_{ori}$. Fig \ref{fig:far_near} compares the CDF plot for error in $\theta_{dir}$ and $\theta_{ori}$. In both cases the error significantly reduces. If we compare the 80th percentile error, $\theta_{dir}$ improves from $8^{\circ}$ to $1.3^{\circ}$ and $phi$ improves from $25^{\circ}$ to $2^{\circ}$. The reason behind this improvement is the more pronounced diversity in the acoustic channels between the ears and the mouth even for a smaller change in $\theta_{dir}$ and $\theta_{ori}$
\new

\begin{figure}[h]
    \centering
    \includegraphics[width=1.6in]{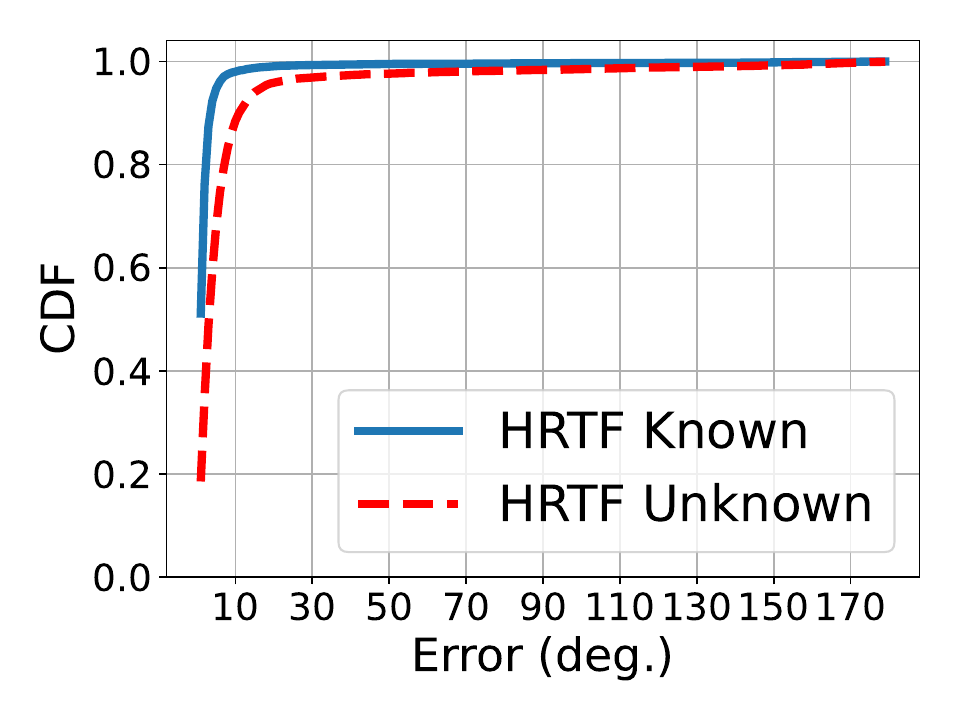}
    \includegraphics[width=1.6in]{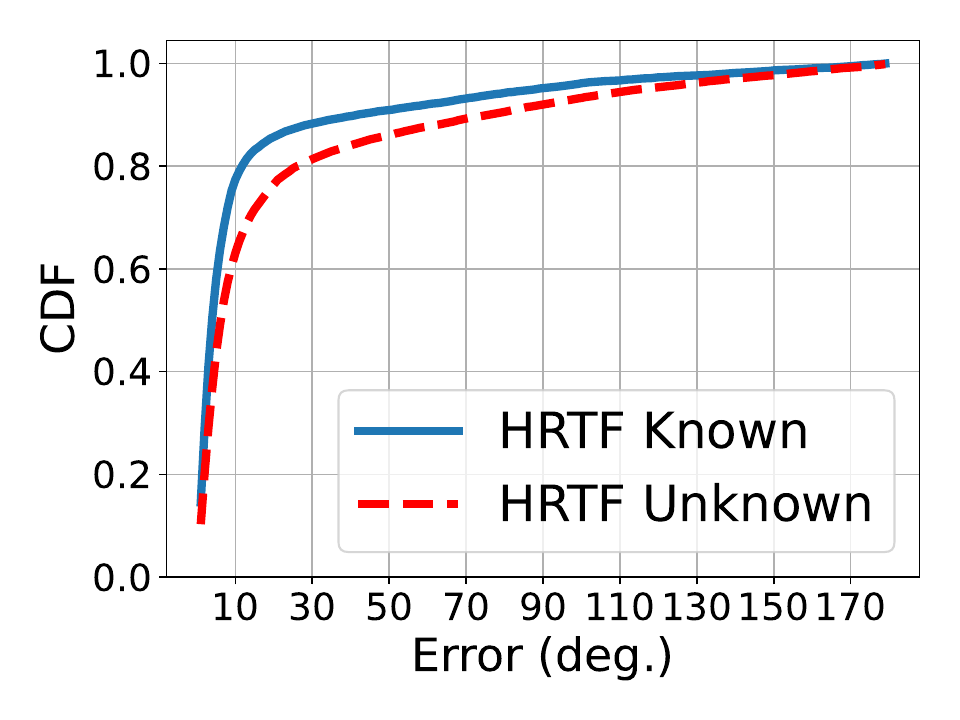}
\vspace{-0.17in}    
\caption{Error in (a) $\theta_{dir}$ and (b) $\theta_{ori}$ for know/unknown HRTF.}
\vspace{-0.2in}
%\caption{Comparison CDF plot for error in estimated (a) listener and (b) speaker direction when HRTF is known vs when HRTF is unknown}
\label{fig:cdf_listener_known_unknown}
\end{figure}

%\noindent{\bf What if the listener HRTF is known (Personalized training):}
\noindent{\bf Performance with personalized training: }
In this section, we compare the results between two cases - 1) when the user's HRTF is known and 2) when it is not known. When the HRTF is known, we re-train our model on that particular HRTF for multiple speakers. Fig \ref{fig:cdf_listener_known_unknown} (a) and \ref{fig:cdf_listener_known_unknown} (b)  shows the comparison between CDF plots for both cases for $\theta_{dir}$ and $\theta_{ori}$ respectively. For both $\theta_{dir}$ and $\theta_{ori}$ the results improve significantly.

%\vspace{-0.1in}

\subsection{Classification performance}
\vspace{-0.1in}
We first evaluate our model for accuracy in an application scenario requiring classification of states to find if the speaker and listener are facing each other. One person is facing if the other is within $25^\circ$ sector in the front. We considered four classes: (a) ``facing \& facing'', (b) ``facing \& non-facing'', (c) ``non-facing \& facing'', and (d) ``non-facing \& non-facing''.
The classifier shows accuracy of $>90\%$ for all configurations except for the ``non-facing \& non-facing'' configuration where the accuracy drops to $89\%$.

\section{Conclusion}
\vspace{-0.1in}
We presented a system for estimating speaker's direction and head orientation using binaural recording on the listener's ear mounted microphones in the near-field which magnifies the diversity and aids estimation. The system leverages the speaker's VDP and listener's HRTF jointly to achieve $90^{th}$ percentile errors of $2.5^{\circ}$ and $12.5^{\circ}$ in direction and orientation respectively.
% We also show that near-field effects can be useful in this scenario to provide additional diversity necessary for head orientation estimation.

% Below is an example of how to insert images. Delete the ``\vspace'' line,
% uncomment the preceding line ``\centerline...'' and replace ``imageX.ps''
% with a suitable PostScript file name.
% -------------------------------------------------------------------------

%\vfill
%\pagebreak

\vfill\pagebreak

\balance
\bibliographystyle{IEEEbib}
\bibliography{strings,refs}
\end{document}